\documentclass[
    twocolumn,
	prd,
    amssymb,
	preprintnumbers,
	secnumarabic,
	nofootinbib,
	superscriptaddress]{revtex4-1}

\pdfoutput=1

\usepackage{graphicx}
\usepackage{enumitem}
\usepackage{latexsym}
\usepackage{amsfonts}
\usepackage{amssymb}
\usepackage{array}
\usepackage{pifont}
\usepackage{color}
\usepackage{xcolor}
\usepackage{amsmath}
\usepackage{slashed}
\usepackage{dcolumn}
\usepackage{verbatim}
\usepackage{float}
\usepackage{multirow}
\usepackage{xspace}
\usepackage[normalem]{ulem}
\usepackage{hyperref}
\usepackage{subfigure}

\definecolor{aquamarine}{rgb}{0.2,0.7,0.6}
\definecolor{hypershade}{rgb}{0.3,0.3,0.8}
\hypersetup{
  pdfauthor={Nirmal Raj},
  pdftitle={Tiny Shiny},
  pdfsubject={Tiny Shiny},
  colorlinks=true,
  citecolor=green,
  urlcolor=purple,
  linkcolor=hypershade
}

\newcommand{\gsim}{\gtrsim}
\newcommand{\lsim}{\lesssim}

\def\Oc{\mathcal{O}}

\newcommand{\beq}{\begin{equation}}
\newcommand{\eeq}{\end{equation}}
\newcommand{\bea}{\begin{eqnarray}}
\newcommand{\eea}{\end{eqnarray}}
\newcommand{\nn}{\nonumber}

\definecolor{rosy}{RGB}{230,235,252}
\definecolor{myframetitle}{RGB}{90,89,170}
\definecolor{myblocktitle}{RGB}{140,185,249}
\definecolor{mytitle}{RGB}{10,80,26}

\definecolor{darkgreen}{RGB}{27,130,45}
\definecolor{darkblue}{rgb}{0,0,0.3}
\definecolor{darkred}{rgb}{0.7,0,0}

\definecolor{light gray}{RGB}{220,220,220}
\definecolor{dark purple}{RGB}{108,0,217}
\definecolor{pink}{RGB}{190,20,100}
\definecolor{orang}{RGB}{193,63,0}
\definecolor{green}{RGB}{11,98,17}
\definecolor{darkpink}{RGB}{153,0,76}
\definecolor{bluegreen}{RGB}{0,102,102}
\definecolor{greenlagan}{RGB}{0,102,0}
\definecolor{redgreen}{RGB}{102,102,0}
\definecolor{Redgreen}{RGB}{153,76,0}
\definecolor{vividviolet}{rgb}{0.62, 0.0, 1.0}
\definecolor{amaranth}{rgb}{0.9, 0.17, 0.31}
\definecolor{palatinateblue}{rgb}{0.15, 0.23, 0.89}
\definecolor{brightpink}{rgb}{1.0, 0.0, 0.5}
\definecolor{cornflowerblue}{rgb}{0.39, 0.58, 0.93}
\definecolor{deepcarminepink}{rgb}{0.94, 0.19, 0.22}
\definecolor{radicalred}{rgb}{1.0, 0.21, 0.37}

\def\TNS{T}
\def\Tinf{T_{\rm \infty}}
\def\MNS{M_{\rm NS}}
\def\RNS{R_{\rm NS}}
\def\Rinf{R_{\rm \infty}}

\allowdisplaybreaks

\begin{document}

\title{Reheated Sub-40000 Kelvin Neutron Stars at the JWST, ELT, and TMT}

\author{Nirmal Raj}
\email{nraj@iisc.ac.in}

\author{Prajwal Shivanna}
\email{prajwals1@iisc.ac.in}

\author{Rachh Gaurav Niraj}
\email{gauravrachh@iisc.ac.in}

\affiliation{Centre for High Energy Physics, Indian Institute of Science, C. V. Raman Avenue, Bengaluru 560012, India}

\date{\today}

\begin{abstract}

Neutron stars cooling passively since their birth may be reheated in their late-stage evolution by a number of possible phenomena: rotochemical, vortex creep, crust cracking, magnetic field decay, or more exotic processes such as removal of neutrons from their Fermi seas (the nucleon Auger effect), baryon number-violating nucleon decay, and accretion of particle dark matter.
Using Exposure Time Calculator tools, we show that reheating mechanisms imparting effective temperatures of 2000--40000 Kelvin may be uncovered with excellent sensitivities at the James Webb Space Telescope (JWST), the Extremely Large Telescope (ELT), and the Thirty Meter Telescope (TMT), with imaging instruments operating from visible-edge to near-infrared. 
With a day of exposure, they could constrain the reheating luminosity of a neutron star up to a distance of 500~pc, within which about $10^5$ (undiscovered) neutron stars lie.
Detection in multiple filters could overconstrain a neutron star's surface temperature, distance from Earth, mass, and radius.
Using publicly available catalogues of newly discovered pulsars at the FAST and CHIME radio telescopes and the Galactic electron distribution models YMW16 and NE2001, we estimate the pulsars' dispersion measure distance from Earth, and find that potentially 30$-$40 of these may be inspected for late-stage reheating within viable exposure times, in addition to a few hundred candidates already present in the ATNF catalogue.
Whereas the coldest neutron star observed (PSR J2144$-$3933) has an upper limit on its effective temperature of about 33000 Kelvin with the Hubble Space Telescope, 
we show that the effective temperature may be constrained down to 20000~Kelvin with JWST-NIRCam, 15000~Kelvin at ELT-MICADO, and 9000~Kelvin with TMT-IRIS.
Campaigns to measure thermal luminosities of old neutron stars would be transformative for astrophysics and fundamental physics. 
\end{abstract}

\maketitle

\section{Introduction}

Neutron stars, celestial objects {\em extraordinaire}, are versatile natural laboratories~\cite{NSfundamental:Nattila:2022evn,BramanteRajCompactDark:2023djs}. 
The cooling of neutron stars (NSs) provides an observational handle on their mass, radius, composition of core, crust and envelope, the equation of state (EoS) of NS matter, and their magnetic fields; data on NS cooling curves has helped to show that nucleons in the NS core exhibit weak superfluidity~\cite{coolingcatalogue:Potekhin:2020ttj}.
While there is much agreement between passive cooling models and observations of NS thermal emission and ages down to $10^{5}$~Kelvin and $10^6$~years, certain NSs anomalously hot for their spin-down age have also been identified~\cite{coolingcatalogue:Potekhin:2020ttj}. 
Meanwhile, the coldest NS on record, PSR J2144$-$3933, has its effective temperature bounded at $<$~33000 Kelvin by Hubble Space Telescope (HST) observations~\cite{coldestNSHST}; its spin-down age is $3\times10^8$~years, implying that if passive cooling were in place its temperature would be a few hundred Kelvin in ``minimal cooling" models~\cite{coolingminimal:Page:2004fy,coolinganalytic:Ofengeim:2017cum,NSvIR:clumps2021}, those that incorporate only modified Urca processes of neutrino cooling. 

In light of these, observational studies of late-time reheating mechanisms of NSs become important.
These mechanisms include astrophysical effects~\cite{NsvIR:otherinternalheatings:Reisenegger} such as rotochemical heating, vortex creep heating, crust-cracking, and magnetic field decay, and those involving a hidden sector of particles~\cite{BramanteRajCompactDark:2023djs} 
such as dark matter capture and heating through scattering and annihilations, 
Bondi accretion of dissipative-fluid dark matter as well as minimal accretion of collisionless dark matter residing in micro-halos, 
accretion through a long-range force between dark matter and neutron stars,
the nucleon Auger effect via neutron oscillations, 
and baryon number-violating neutron decays; see Appendix~\ref{app:reheat} for more discussion on these mechanisms.\footnote{Despite the name ``reheating", these mechanisms do not heat the NS more than once. 
They continuously deposit energy in the NS material over a long time, their effects only becoming apparent when the NS cools down to a luminosity comparable to the energy deposition rate.}
It is timely that the currently operational James Webb Space Telescope (JWST)~\cite{JWST:Gardner:2006ky} that is delivering a raft of discoveries~\cite{JWSTDiscoveries},
the soon-to-appear Extremely Large Telescope (ELT)~\cite{ELT:neichel2018overview},
and the less imminent Thirty Meter Telescope (TMT)~\cite{TMT:2015pvw} can span wavelengths corresponding to blackbody peak temperatures of 1300--4300 Kelvin with their imaging instruments, respectively 
the Near Infrared Camera (NIRCam),
the Multi-AO Imaging Camera for Deep Observations (MICADO), and
the InfraRed Imaging Spectrograph (IRIS).
That is, all three telescopes can cover from the far-optical down to the infrared.

In this study we estimate the sensitivities of these telescopes to NS temperatures in this range.
We find that $> 10^5$ pulsars may be potentially measured at a reheated temperature of $\lsim$~40000 Kelvin. 
(As the distribution of pulsar ages $t_{\rm NS}$ is roughly weighted as $t_{\rm NS}/t_{\rm MW}$~\cite{NSdistribs:Sartore2010}, where $t_{\rm MW} = 5$~Gyr is the age of the Milky Way, we expect $>1000$~K pulsars merely cooling passively, i.e. pulsars with $t_{\rm NS} < 10^7$~yr, to be rare. 
Hence our emphasis on potentially reheated neutron stars.)
Statistically significant measurements of the spectral flux density of thermal emission of an NS in multiple filters could help to simultaneously pin down the NS surface temperature, distance, mass, and radius.
To launch a campaign of observing NS reheating, some fraction of this NS population must be discovered as pulsars in order to train the infrared and optical telescopes at the right sky position.
However, the ATNF catalogue~\cite{ATNF:2004bp} already contains a few hundred pulsar candidates close enough to be amenable to luminosity measurements. 
Further, in this work we avail of publicly available catalogues of radio pulsars discovered in the currently operational Five-hundred-meter Aperture Spherical Telescope (FAST) and Canadian Hydrogen Intensity Mapping Experiment (CHIME) to extract their (dispersion measure-based) distances from Earth, showing that a few tens of freshly spotted pulsars are close enough to be good targets for observations of thermal emission at JWST, ELT, and TMT.

Moreover, we argue that there is already a clear observational target for these telescopes to improve our understanding of the NS cooling and reheating, namely, the coldest NS observed, PSR J2144$-$3933, that was mentioned earlier.
While an upper limit on its thermal luminosity exists thanks to HST, this limit can be improved by a factor of 7 by JWST-NIRCam, a factor of 20 by ELT-MICADO, and a factor of 180 by TMT-IRIS.
Thus there is immediate opportunity to push the ``neutron star temperature frontier" and revise constraints on NS reheating mechanisms offered by astrophysics and particle physics.

 Yet another campaign may be mounted: the observation of (already discovered) X-ray sources that are almost certainly neutron stars. 
These are ``central compact objects" (CCOs), located near the centers of supernova remnants that lack pulsar wind nebulae~\cite{DeLucaCCOs:2017rdc}.
As described in detail in Sec.~\ref{sec:disc}, CCOs could be surrounded by a disk of debris left behind by the neutron star's progenitor after supernova explosion.
This disk could source an infrared flux that is obtained in addition to the X-ray thermal peak.
A search for such disk emission has already been undertaken recently using JWST-NIRCam in the Cas A system, yielding a null result~\cite{CasAJWSTMilisavljevic:2024mbg}.
At the time of writing, there are nine other confirmed CCOs~\cite{CatalogueCCODeLuca} for which similar searches can be performed.

This document is organized as follows.
In Section~\ref{sec:signals} we set up the basic framework of imaging sensitivities, 
describe the tools that estimate exposure times for our telescopes and our choices of input parameters,
and present our results.
We also discuss the future reach of these telescopes on the temperature of PSR J2144$-$3933.
In Section~\ref{sec:disc} we provide some discussion and conclude.
In the appendices we review various late-stage reheating mechanisms in the literature, 
and tabulate our derived dispersion measure distances of pulsars in recently available catalogues made public by FAST and CHIME.

\begin{figure*}
    \centering
    \includegraphics[width=0.47\textwidth]{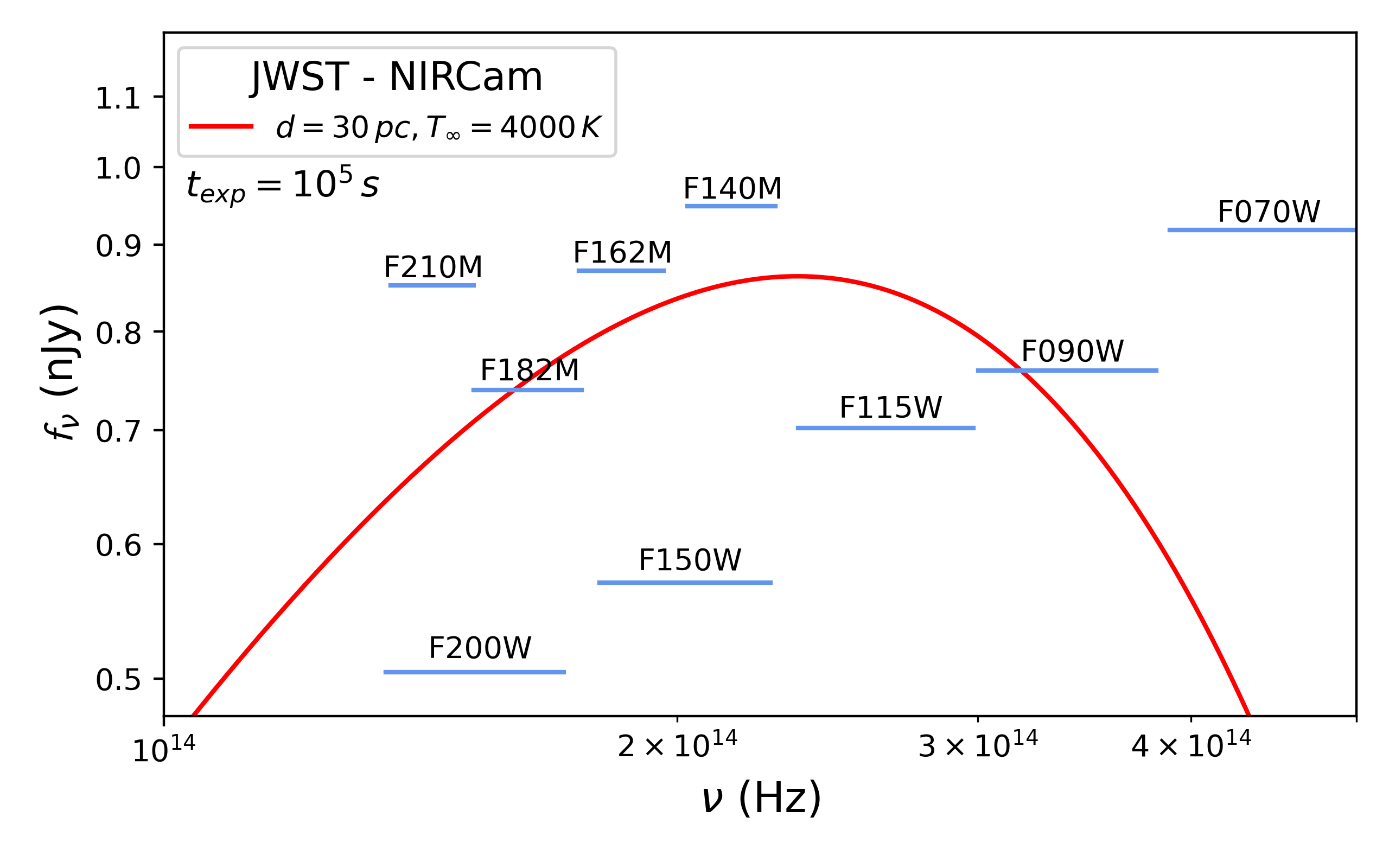} \     \includegraphics[width=0.47\textwidth]{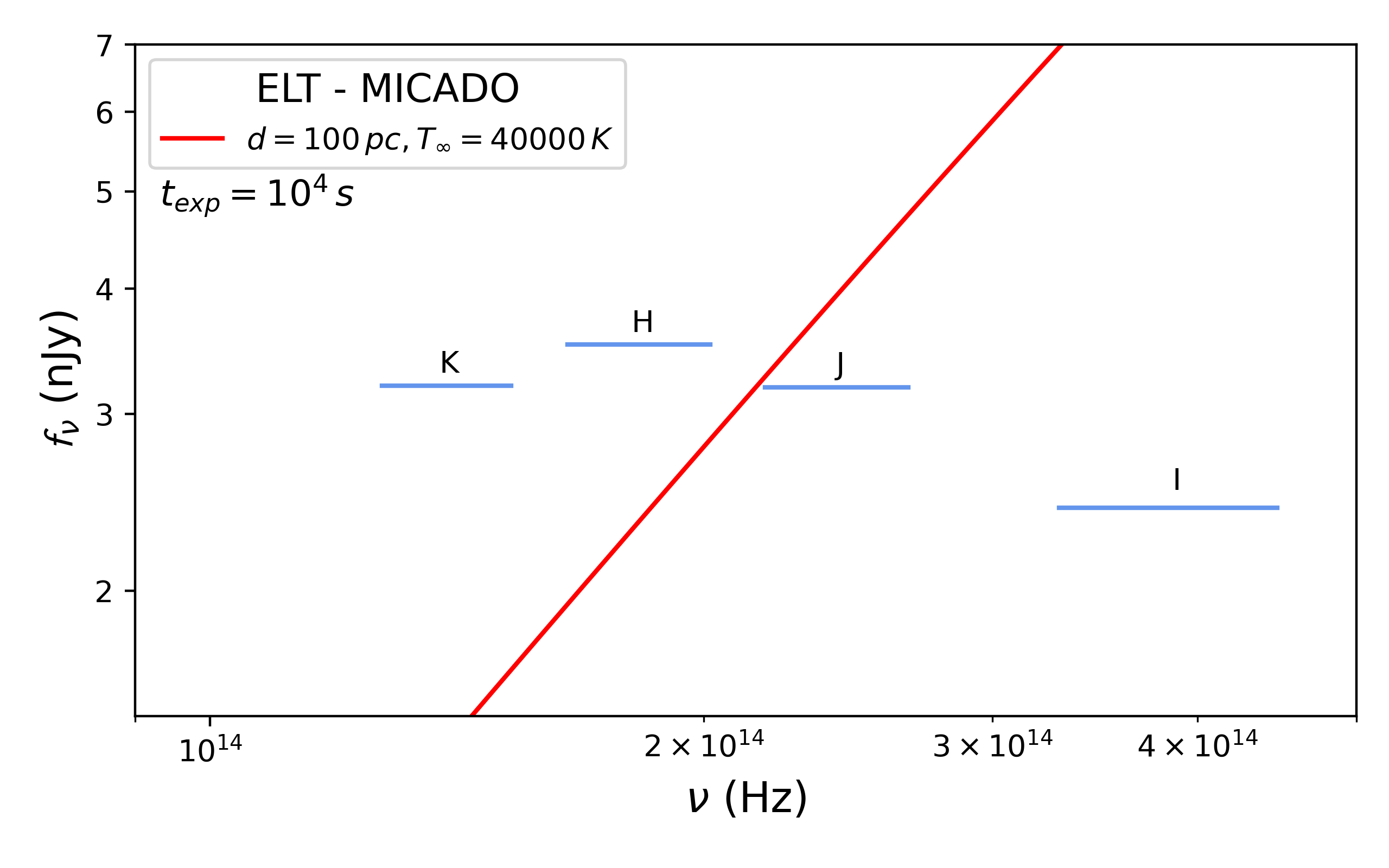}
    \caption{Blackbody spectral flux densities of neutron stars with various effective temperatures at various distances compared with filter sensitivities at JWST-NIRCam and ELT-MICADO at SNR=2 for the indicated exposure times.
     Luminosity measurements of late-time-reheated neutron stars at these imaging instruments are eminently feasible. 
     See Sec.~\ref{sec:signals} for further details.}
    \label{fig:fnu}
\end{figure*}

\begin{figure*}
    \centering
          \includegraphics[width=0.95\textwidth]{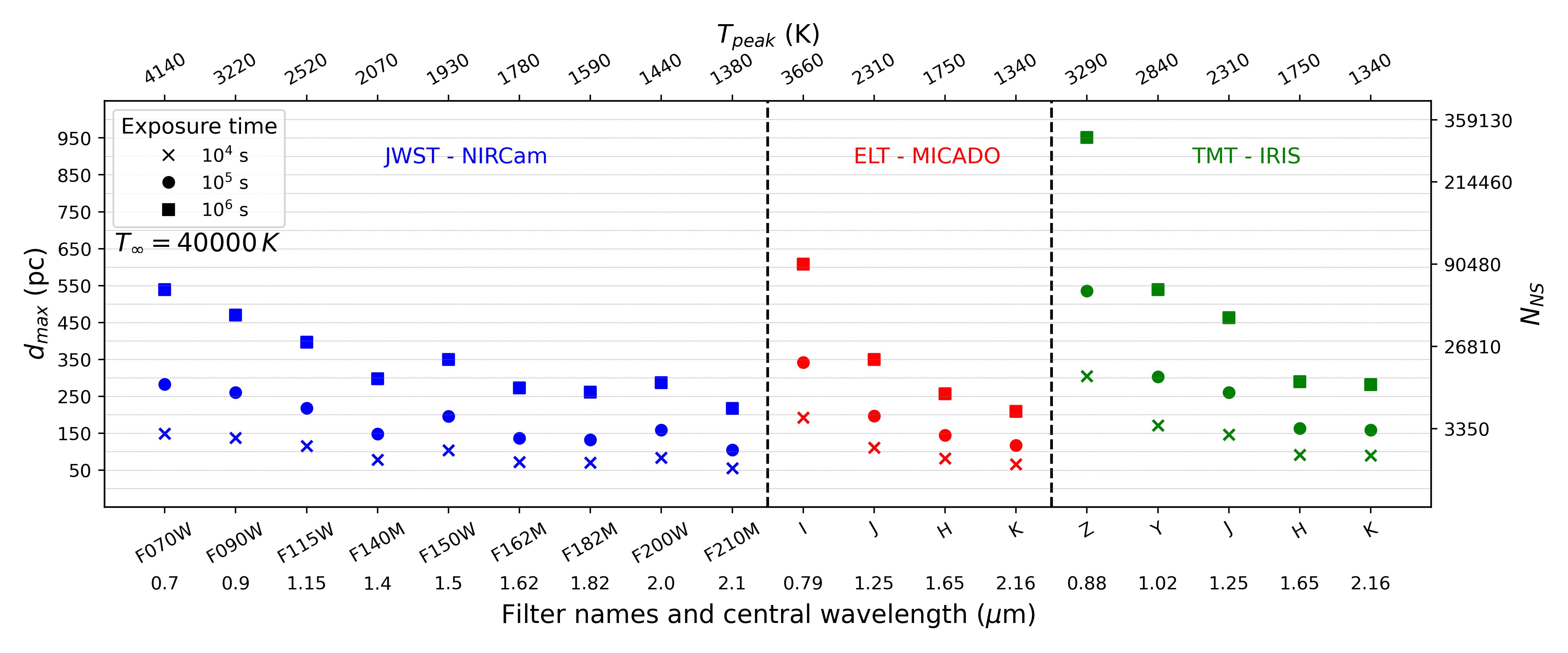} \\
           \includegraphics[width=0.95\textwidth]{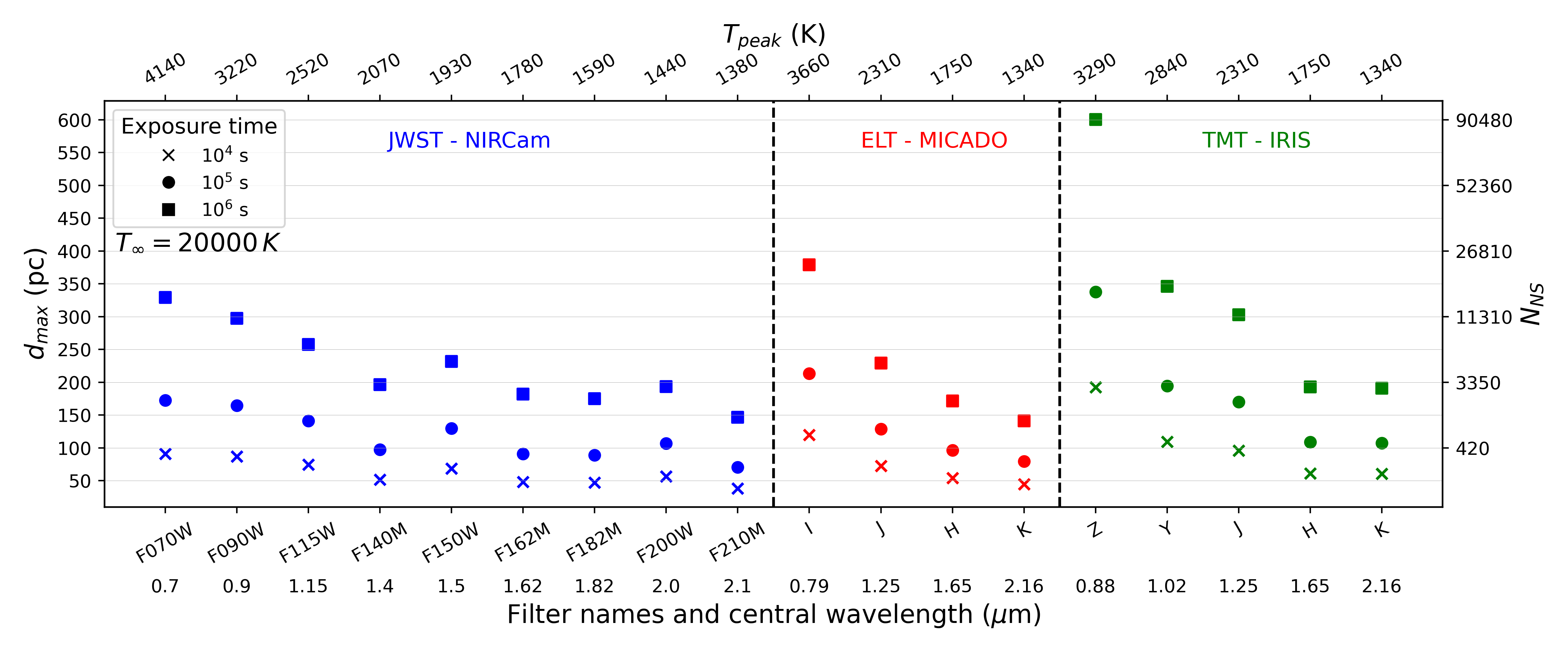} \\
           \includegraphics[width=0.95\textwidth]{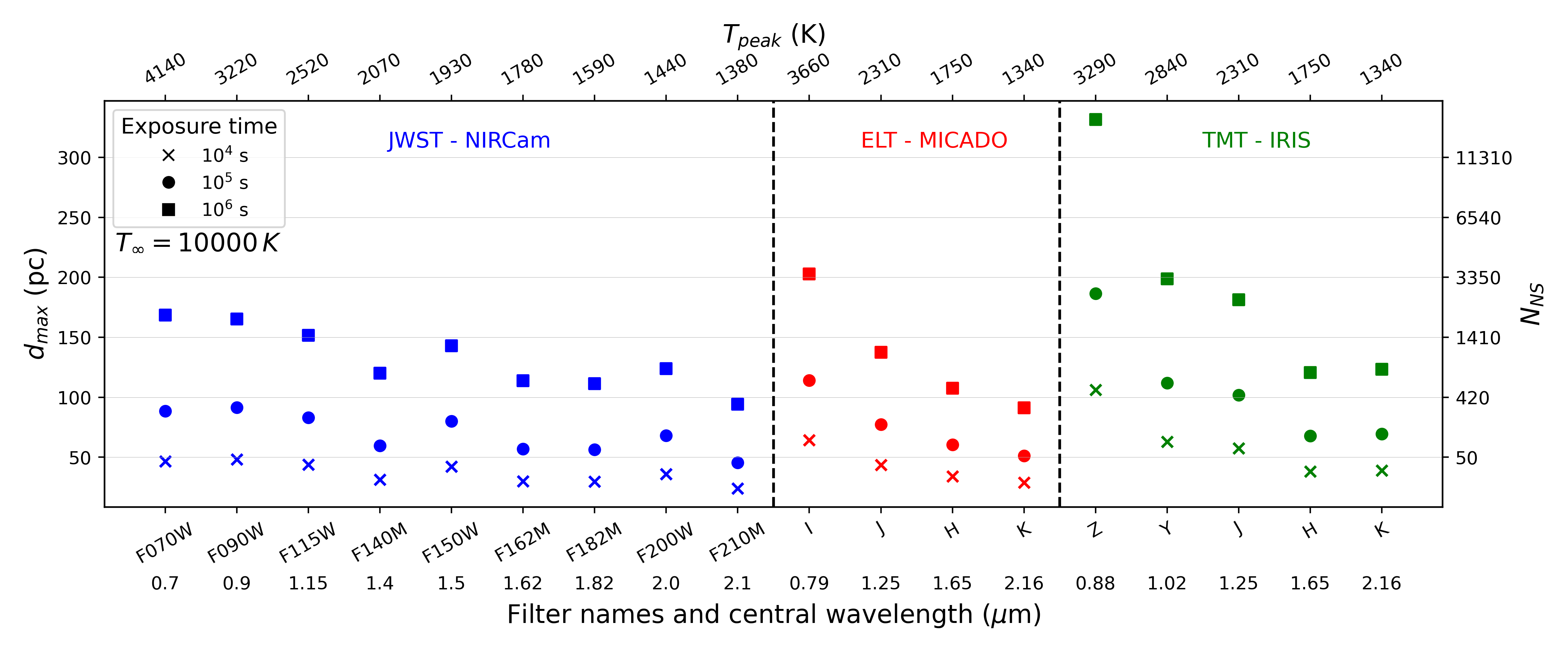} 
     \caption{The maximum distance to which neutron stars with effective temperatures of 40000~K, 20000~K and 10000~K can be constrained at SNR 2 in the filters of JWST-NIRCam, ELT-MICADO and TMT-IRIS, with net exposure times of $10^4$, $10^5$, $10^6$~seconds. 
     These temperatures are within an order of magnitude of the upper limit on the effective temperature of the coldest neutron star on record (PSR J2144$-$3933), about 33000~K, and are also above the blackbody peak temperatures corresponding to the central wavelength of the filters here. 
 Not shown for visual clarity are astrophysical modelling uncertainties at the 10\% level, comparable to uncertainties in distance measurements via radio parallax and much smaller than uncertainties in distance estimates via pulsar dispersion measure.
 The plot data are tabulated and uploaded to arXiv as an ancillary file.
 See Sec.~\ref{sec:signals} for further details.}
    \label{fig:dmax}
\end{figure*}


\begin{figure*}
    \centering
    \includegraphics[width=0.95\textwidth]{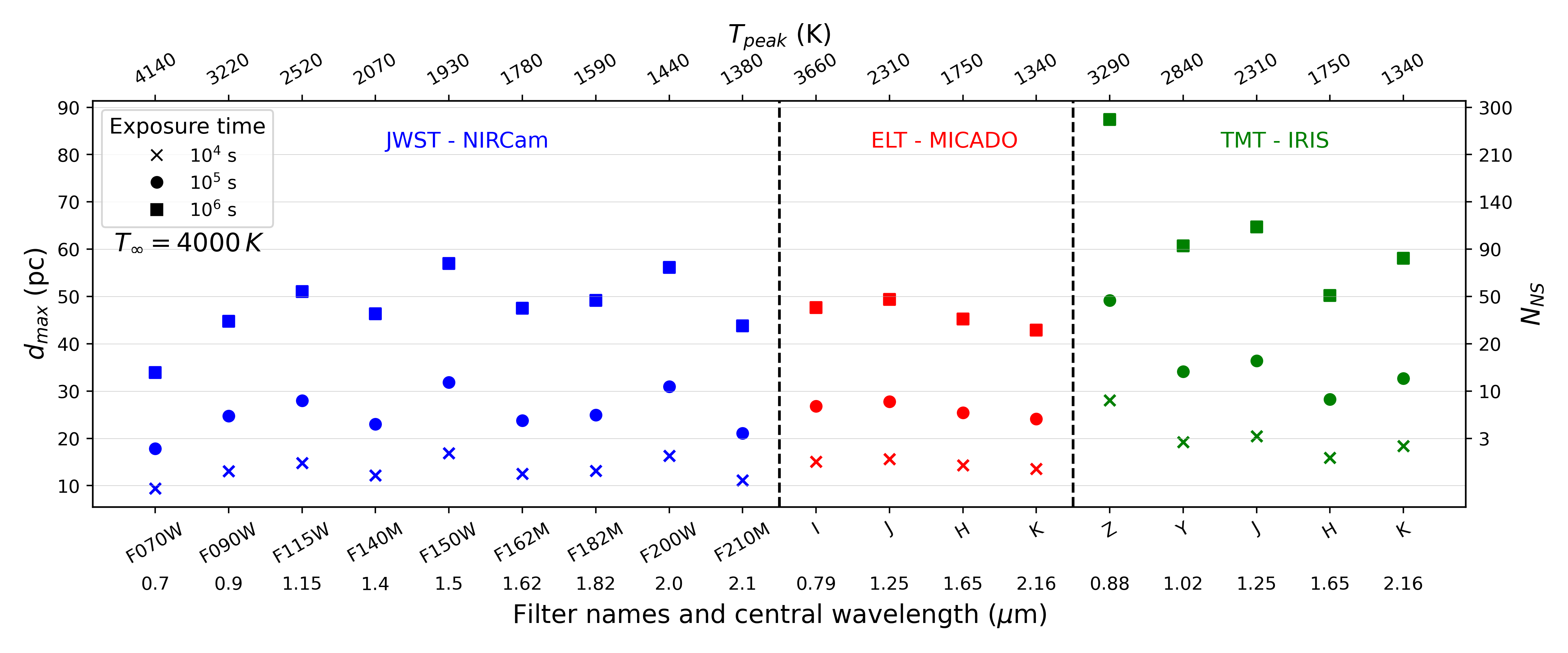} \\
   \includegraphics[width=0.95\textwidth]{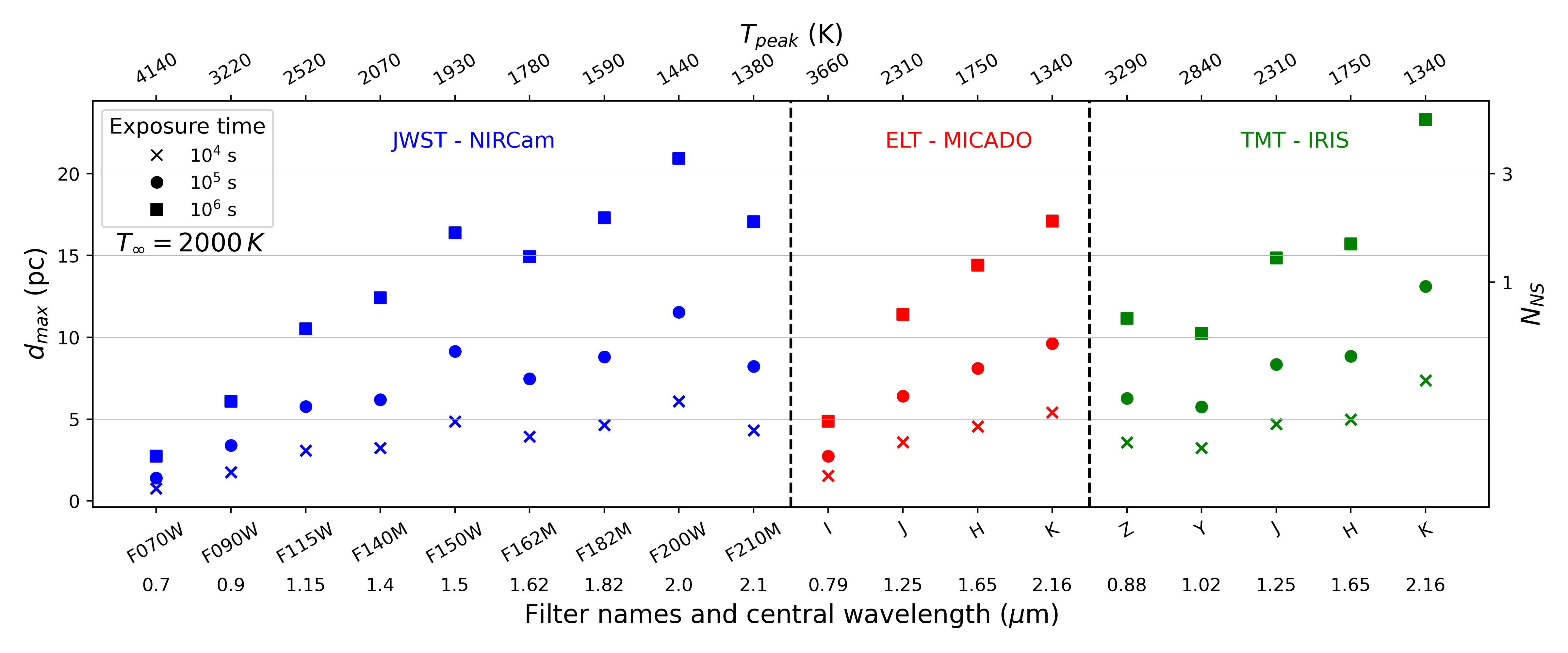} 
    \caption{Same as Fig.~\ref{fig:dmax} but for effective temperatures of 4000~K and 2000~K. These temperatures are below the blackbody peak temperatures corresponding to the central wavelength of the filters here. See Sec.~\ref{sec:signals} for further details.}
    \label{fig:dmax2}
\end{figure*}

\section{Signals and Results}
\label{sec:signals}

\subsection{General set-up}
\label{subsec:gensetup}

The blackbody spectral flux density of an NS of mass $\MNS$, radius $\RNS$, surface temperature $T$ (so that the radius and effective temperature as seen by a distant observer are $\Rinf =  (1+z) \RNS$ and $\Tinf = \TNS/(1+z)$, where $1+z = 1/\sqrt{1-2G\MNS/\RNS}$ )  at a distance from Earth $d$ is given by
\bea
\label{eq:fnu}
f_\nu &=& \pi B (\nu, \Tinf) \bigg(\frac{\Rinf}{d}\bigg)^2~\times 10^{-0.4 A_\nu}, \\
\nn {\rm with} \ B_\nu &=& \frac{2 h \nu^3}{c^2} \frac{1}{\exp(h\nu/k\Tinf)-1}~,
\eea
where $A_\nu$ is the extinction factor determined by the column density of Galactic dust along the line of sight to the NS.
Extinction will modify our results at worst by about 10\%, which is within the uncertainties of measurements of NS distances. 
Therefore we show our main results by setting $A_\nu = 0$, and discuss the effect of extinction in Sec.~\ref{subsec:trends}.

The signal-to-noise ratio (SNR) as a function of exposure time $t_{\rm exp}$ roughly goes as $F_{\rm sig} A_{\rm SNR} t_{\rm exp}/\sqrt{(F_{\rm bg}A_{\rm SNR}+\Gamma_{\rm noise})t_{\rm exp}}$, where $A_{\rm SNR}$ is the SNR reference area in the detector, $F_{\rm sig}$ and $F_{\rm bg}$ are signal and background fluxes, and $\Gamma_{\rm noise}$ is the rate of noise from non-sky sources.
Thus $t_{\rm exp} \propto$~(SNR)$^2$.
As we will see soon, further subtleties enter the calculation of exposure time, but the above expression helps us understand some salient features.
It tells us that for a given $\Tinf$, there is a maximum distance $d_{\rm max}$ up to which NSs can be observed in a given instrument filter that comes with a flux sensitivity corresponding to some SNR and exposure time.
For a solar neighborhood NS number density $n_{\rm NS, \odot}$, there is a maximum of $N_{\rm max} = 4\pi n_{\rm NS} d^3_{\rm max}/3$ potential NS targets.
From Ref.~\cite{NSdistribs:Sartore2010}, $n_{\rm NS, \odot} \simeq (1-5) \times 10^{-4} (N_{\rm NS}/10^9)~{\rm pc}^{-3}$, the factor of 5 variation arising from modelling of the local NS population. 
The Milky Way NS population $N_{\rm NS}$ could be $10^8$ for a constant supernova rate, however it could be as high as $10^9$ via a higher past supernova rate as suggested by Galactic nucleosynthesis constraints~\cite{NSbirthratehigh:Arnett:1988sn}.
Given these uncertainties, we take $n_{\rm NS, \odot} = 10^{-4}~{\rm pc}^{-3}$ as a representative value to show our results.

In Fig.~\ref{fig:fnu} we plot $f_\nu$ for $\Tinf = 4000$~K, $d = 30$~pc (left panel) and $40000$~K, $d = 100$~pc (right panel), and show for comparison the SNR = 2 sensitivities of filters at JWST-NIRCam (for $t_{\rm exp} = 10^4$~s) and ELT-MICADO (for $t_{\rm exp} =  10^5$~s).
This figure illustrates that observations of thermal emission of NSs of temperatures and distances considered here can indeed be made with the right choice of filters.
In Figs.~\ref{fig:dmax} and Fig.~\ref{fig:dmax2}, we visually tabulate $d_{\rm max}$ -- and the corresponding $N_{\rm max}$ on the right-hand y-axis -- against various filters, indicating their central wavelengths in the lower x-axis and the corresponding peak blackbody temperatures in the upper x-axis.
We do this for an SNR of 2 (corresponding to a 95\% C.L. limit in Gaussian statistics) and total exposure times of $10^4, 10^5, 10^6$ seconds.
The first of these choices ($\sim \Oc$(hours)) is a reasonable telescope time to request, as seen in several JWST proposals~\cite{JWSTProposals}.
The last choice ($\sim \Oc$(months)) is achievable in deep exposures as undertaken by HST.
The intermediary exposure timescale (about a day) has been quoted as a viable exposure time at JWST in the literature~\cite{NSvIR:IISc2022}.
The panels of Figs.~\ref{fig:dmax} and \ref{fig:dmax2} correspond to different $\Tinf$ as indicated.
The case of  $\Tinf = 40000$ Kelvin is near the observational upper limit on NS temperatures~\cite{coldestNSHST}, and is hence inherently interesting.
The cases of lower temperatures are those corresponding to reheating mechanisms in the literature discussed in the Introduction and Appendix~\ref{app:reheat}.
To show our results we had assumed a benchmark NS mass and radius of 
\beq
M_{\rm NS} = 1.4~M_\odot~,~~\RNS = 11~{\rm km}~.
\label{eq:NSBM}
\eeq
These values are typical of NSs and were assumed by Ref.~\cite{coldestNSHST} to illustrate the temperature measurement of PSR J2144$-$3933. 
(Note that, with this choice, the limit on the redshifted temperature is $\Tinf <$ 33000 K.)
These values also happen to fall in the mass-radius curve predicted by the EoS of Wiringa, Fiks, Fabrocini (WFF)~\cite{EoS:Wiringa:1988tp,crustdensityprofileRRIIIA}.
We will discuss in Sec.~\ref{subsec:trends} the effect of varying the NS mass and radius along the WFF EoS prediction.
\begin{table*}[]
    \centering
    \begin{tabular}{c c c c}
      \multicolumn{4}{c}{\bf JWST/NIRCam, DEEP8 readout} \\
    \hline
     net exposure time ($s$)  & dithers & integrations per exposure & groups per integration \\
     \hline
      $10^4$  & 17 & 2 & 2 \\ 
       10$^5$  &  21 & 21 & 2 \\
       10$^6$ & 21 & 22 & 11 \\
    \hline
    \end{tabular}
    \caption{Our choice of parameters to achieve desired net exposure times in the JWST Exposure Time Calculator tool. See Sec.~\ref{subsec:jwstnircam} for further particulars.}
    \label{tab:JWSTETC}
\end{table*}

\subsection{JWST-NIRCam}
\label{subsec:jwstnircam}

Our NIRCam results are obtained using the online tool JWST Exposure Time Calculator (ETC) version 3.0.
Detailed documentation of the working of the ETC is available~\cite{JWSTExpoTimes}, and here we summarize relevant points.
In the ETC one must make an optimal choice from among 9 available readout patterns, which fix the number of usable detector readout samples (``frames") in a ``group" of them.
For the full detector, the sampling time $t_{\rm frame}$ is 10.737~s.
A collection of groups is an ``integration", and a collection of integrations is an ``exposure".
To achieve a net exposure time, the ETC user can tune the groups per integration $N_{\rm group}$ and the integrations per exposure $N_{\rm int}$, but not the frames per group, which is fixed by the readout pattern.
Within a readout pattern, a certain number of frames are averaged for signal-fitting in a group ($N_{\rm frame}$), while others are skipped ($N_{\rm skip}$).
Skipped samples at the end of an integration are not processed, so the integration time $t_{\rm int}$ is given by 
\beq
t_{\rm int} = t_{\rm frame} [(N_{\rm group } - 1) (N_{\rm frame} + N_{\rm skip}) + N_{\rm frame}]~.
\label{eq:tint}
\eeq
For faint sources such as those we are interested in, increasing $N_{\rm int}$ while decreasing $N_{\rm group}$ is the preferred strategy, to reduce the data volume that is downlinked to the ground.
Finally, the user must suitably choose the number of ``dithers", small shifts in telescope pointing made between exposures to improve sky coverage and image quality by combining resultant images and thereby removing detector artifacts and cosmic ray hits, filling inter-detector gaps, and calibrating the point spread function. 

In Table~\ref{tab:JWSTETC} we show our choices of the number of dithers, $N_{\rm int}$ and $N_{\rm group}$ to achieve our net exposure times. 
As appropriate for long integration times, we used {\tt DEEP8}, the readout pattern with the maximum number of frames averaged in a group (which is 8, with 12 other frames skipped) and with the maximum number of groups allowed.
Such a readout pattern is susceptible to pixel loss due to cosmic ray impacts, hence we select a high number of dithers.
As in Ref.~\cite{NSvIR:IISc2022}, we picked a reference background model at RA=$03^{\rm h} \ 32^{\rm m} \ 42.397^{\rm s}$ and Dec= $-27^\circ \ 42' \ 7.93''$, corresponding to a blank field. 
Due to how the data volume is handled and the use of dithering, the SNR grows with exposure time at a rate slightly smaller than $\sqrt{t_{\rm exp}}$.

{\bf NIRISS sensitivities.}

In addition to NIRCam, JWST can perform imaging with the Near-Infrared Imager and Slitless Spectrograph (NIRISS) instrument. 
The sensitivities of NIRISS' filters are generally slightly weaker than NIRCam's and for this reason we do not display NIRISS results.
By comparing NIRCam and NIRISS filter sensitivities in the JWST Pocket Guide~\cite{JWSTPocketGuide} and by using Eq.~\eqref{eq:fnu}, we find that the ratios of $d_{\rm max}$ obtained in NIRISS versus NIRCam in various filters are
{F090W: 0.87, F115W: 0.89, F140M: 0.89, F150W: 0.84, F158M/F162M: 0.88, F200W: 0.85, F277W: 1.07,
F356W: 0.99, F430W: 0.90, F444W: 0.93, F480M: 0.90}.
In the JWST Pocket Guide the NIRCam sensitivities  are based on the {\tt MEDIUM8} readout pattern with 10 exposures and $N_{\rm group} = 10$, while the readout pattern of NIRISS sensitivities is not specified but is presumably {\tt NIS}~\cite{NIRISSReadout} as appropriate for faint sources.
If the {\tt DEEP8} readout pattern were used at NIRCam instead (as done in our study), the NIRCam sensitivities would be even better than those of NIRISS.

\subsection{ELT-MICADO}
\label{subsec:eltmicado}

For the terrestrial telescopes ELT and TMT imaging through the atmosphere, other considerations come into play.
As per the ELT ETC documentation~\cite{ELTETC}, the signal-to-noise ratio is given by
\beq
{\rm SNR} = \frac{N_{\rm exp}N_{\rm obj}}{\sqrt{N_{\rm exp}[N_{\rm obj}+N_{\rm sky}+N_{\rm pix}(N_{\rm ro}^2 + \Gamma_{\rm DC} t_1 )]}}~,
\eeq
where $N_{\rm exp}$ is the number of exposures, 
$N_{\rm pix}$ is the number of pixels in the SNR reference area,
$N_{\rm ro}$ is the detector read-out noise (in $e^-$/pixel), 
$\Gamma_{\rm DC}$ is the detector dark current (in $e^-$/s/pixel), 
and $t_1$ is the detector exposure time for one exposure.
The number of detected electrons per exposure from the source in the SNR reference area is
$N_{\rm obj} = \hat \Phi_{\rm obj} \varepsilon  \Omega \alpha_c t_1$, where $\hat \Phi_{\rm obj} =  10^{-0.4 \chi k} \Phi_{\rm obj}$ is the source flux at the telescope entrance with $\Phi_{\rm obj}$ the source flux at the top of the atmosphere and the pre-factor the atmospheric extinction characterized by the airmass $\chi$ and the extinction coefficient $k$,
$\varepsilon$ is the ensquared energy in the SNR reference area as tabulated in the ETC documentation,
$\alpha_c$ is the conversion factor from energy flux at the telescope entrance to detected number of photo-electrons,
and $\Omega = 10^{-6} N_{\rm pix} p_{\rm sc}^2$ is the size of the SNR reference area (in arcsec$^2$);
$p_{\rm sc}$ is the pixel scale (in mas/pixel).
The number of detected electrons per
exposure from the background in the SNR reference area is $N_{\rm sky} = \hat \Phi_{\rm sky} \Omega \alpha_c t_1$, with $\hat \Phi_{\rm sky}$
the sky surface brightness at the telescope entrance.

In the ELT ETC (version 6.4.0) we chose the observatory site as ``Paranal", which has an altitude of 2635 m that is close to that of the actual proposed site, Cerro Armazones at 3046 m.
We also set the telescope diameter to 39 m as currently designed, the adaptive optics (AO) mode to multi-conjugate AO as available on ELT's near-infrared imaging instrument MICADO, and the airmass $\chi$ to 1.5. 
MICADO has fields of view $50.5'' \times 50.5''$ with pixel scale 4 mas/pixel and $19'' \times 19''$ with pixel scale 1.5 mas/pixel, both implying 160 pixels each~\cite{ELTScience:Padovani:2023dxc}.
Thus we set the SNR reference area to 10$\times$10 pixels and pixel scale to 5 mas/pixel, the closest available options.

\subsection{TMT-IRIS}
\label{subsec:tmt-iris}
In the Exposure Time Calculator of TMT's versatile instrument IRIS~\cite{TMTIRISETC},
we take the default zenith angle of observation of 30 degrees,
the default ``average" atmospheric conditions,
and the PSF location at ($8.8'', 8.8''$), corresponding to the centre of the detector field of view.
The imager's field of view is $34'' \times 34''$ with pixel scale 4 mas/pixel.

\subsection{Discussion of results}
\label{subsec:trends}

As mentioned in Sec.~\ref{subsec:gensetup} for deep exposures the SNR~$\propto \sqrt{t_{\rm exp}}$ approximately, and thus from Eq.~\eqref{eq:fnu}, $d_{\rm max} \propto t_{\rm exp}^{1/4}$. 
This is just what we see in Figs.~\ref{fig:dmax} and \ref{fig:dmax2} as we scan vertically up for each filter.
In Fig.~\ref{fig:dmax}, where $T_\infty > 4000$~K, the $d_{\rm max}$ sensitivity decreases with increasing filter central wavelength more or less monotonically, as expected for spectral flux densities falling with wavelength (and as seen in Fig.~\ref{fig:fnu}).
Deviations from monotonic decrease are due to differences in filter sensitivities.
For $T_\infty = 4000$~K in Fig.~\ref{fig:dmax2}, the peak $d_{\rm max}$ reach for each instrument is obtained for one of the intermediary filters with central wavelength corresponding to peak blackbody temperature somewhat below 4000 K.
This is also not surprising as the best result obtainable is an outcome of interplay between the location of the spectral peak and filter sensitivities.
For $T_\infty = 2000$~K in Fig.~\ref{fig:dmax2} the $d_{\rm max}$ reach for each instrument is seen to roughly increase with the central wavelength of the filter choice, as the low-wavelength filters now catch the low wavelength tail of the NS's blackbody spectrum.

Across instruments, it may seen that for comparable filter ranges ELT-MICADO is somewhat more sensitive than JWST-NIRCam, and TMT-IRIS generally yet more sensitive.
In fact for large $T_\infty$ NSs the $Z$ filter of TMT-IRIS gives conspicuously better reach than all other filters considered.

\begin{table*}[t]
   \centering
    \begin{tabular}{ccccc}
      \multicolumn{5}{c}{{\bf PSR J2144$-$3933 effective temperature sensitivities in Kelvin}} \\
      \hline
        \textbf{distance (pc)} & \textbf{exposure (s)} & \textbf{JWST-NIRCam: F707W} & \textbf{ELT-MICADO: {\em I} band} & \textbf{TMT-IRIS: {\em Z} band} \\ \hline
        \multirow{2}{*}{157} & $10^4$ & \underline{43450} {\em 95130} & \underline{29240} {\em 61350} & \underline{15370} {\em 29070} \\ 
        & $10^5$ & \underline{17880} {\em 32940} & \underline{13820} {\em 24630} & \underline{8510} {\em 13350} \\ \hline
        \multirow{2}{*}{172} &$10^4$ & \underline{50150} {\em 112350} &  \underline{33580} {\em 71970} & \underline{17240} {\em 33500} \\ 
        & $10^5$ & \underline{19910} {\em 37550} &  \underline{15330} {\em 28100}  &  \underline{9250} {\em 14860} \\ \hline
        \multirow{2}{*}{192} & $10^4$ & \underline{60670} {\em 137240} &  \underline{39940} {\em 87600} &  \underline{19940} {\em 39990} \\ 
        & $10^5$  & \underline{23350} {\em 45360} &  \underline{17510} {\em 33160} &  \underline{10290} {\em 17050} \\ \hline
    \end{tabular}
 \caption{The SNR = 2 (\underline{underlined}) and SNR = 5 ({\em italicized}) sensitivities, for two practically feasible exposure times 10$^4$ and 10$^5$~s, of the effective temperature in Kelvin of the PSR J2144$-$3933, the coldest neutron star known with a current upper bound on its effective temperature of 33000~Kelvin set by non-detection at HST~\cite{coldestNSHST}. 
 Here a 1.4~$M_\odot$ mass and 11 km radius are assumed.
 Three pulsar parallax distances as estimated in Ref.~\cite{Deller2009:J2144dist} are considered: the central one, and the ends of the $\pm 1\sigma$ range.
 We see that JWST, ELT and TMT are capable of pushing the neutron star temperature frontier lower, potentially improving limits on late-time reheating mechanisms.
 See Sec.~\ref{subsec:J2144} for further particulars. }
    \label{tab:J2144-3933}
    \end{table*}

One may wonder if there is a {\em minimum} NS distance to which the imaging filters here are sensitive.
An NS that is too close may be so bright that it saturates the detector pixels within an integration time.
We checked for this effect in the JWST ETC, and found that saturation occurs if we artificially increase the spectral flux density by 4$-$5 orders of magnitude.
From Eq.~\eqref{eq:fnu}, this implies that NS distances smaller by an $\Oc(100)$ factor than the $d_{\rm max}$ values shown in Figs.~\ref{fig:dmax} and \ref{fig:dmax2} would result in detector saturation.
As even the closest NS is theorized to be about 10 pc from Earth, and the closest observed is about 100 pc away, we may conclude that our results apply to all known NSs whose thermal luminosities are unmeasured. 

While we have used the benchmark NS in Eq.~\eqref{eq:NSBM}, other mass-radius configurations will introduce only $\Oc(10\%)$ variation in the spectral flux density, hence in $d_{\rm max}$.
For example, if we consider a low-mass configuration on the mass-radius curve predicted by the WFF EoS, with $\MNS = 0.95~M_\odot$ and $\RNS = 11$~km, the spectral flux density in Eq.~\eqref{eq:fnu} reduces by 0.92 and hence $d_{\rm max}$ increases by 1.04.
For a high-mass configuration with $\MNS = 2~M_\odot$ and $\RNS = 10.6$~km, $f_\nu$ increases by 1.18 and hence $d_{\rm max}$ reduces by 0.92.
We had also set the interstellar extinction factor $A_\nu$ to zero, but since  $d_{\rm max} \propto 10^{-0.2 A_\nu}$ as obtained from Eq.~\eqref{eq:fnu}, and since typically $A_\nu \ll 1$, our estimate of $d_{\rm max}$ will again only reduce by a small factor if extinction is present along the line of sight.
The average Milky Way extinction factor in the $V$ band (centred at 0.55 $\mu$m) is $A_V = 0.0186$ as derived from the average visual extinction-to-reddening ratio of 3.1~\cite{extinctionreddening1975}, with smaller $A_\nu$ predicted for larger wavelengths by extinction curves~\cite{extinctioncurvesGordon:2003ak} since scattering by interstellar grains becomes weaker. 
Setting our $A_\nu$ even to this maximal extinction, $d_{\rm max}$ decreases but by a factor of 0.92.
These uncertainties are comparable to those of pulsar distance measurements from radio parallax, which achieve a precision of 10-20\%~\cite{Verbiest2010LutzKelkerPulsar,Verbiest2012:pulsardists}.
Where parallax data is unavailable, the radio dispersion measure is used to estimate neutron star distances, which suffer from much larger uncertainties as we will discuss in Sec.~\ref{sec:disc}.

Now suppose an NS is observed to shine with detectably high spectral flux density in some filter. 
Due to uncertainties in its distance from Earth, its mass and radius, the NS' luminosity and temperature cannot be reliably inferred.
However a follow-up observation in a second filter (perhaps nearby in wavelength range) can immediately take the distance (and the corresponding uncertainty) out of the equation if the ratio of $f_\nu$ in two filters is taken, as seen from Eq.~\eqref{eq:fnu}.
This would constrain the temperature to a narrow range, which in fact is how temperatures are inferred from the ``color", i.e. the difference in magnitudes obtained in two filters, as in a color-magnitude diagram.
In fact observations in four different filters would help simultaneously fit $\Tinf$, $d$, $\MNS$ and $\RNS$ in Eq.~\eqref{eq:fnu}, and even more filters could confirm and improve the fit, or in other words, overconstrain the parameters.
The precision achievable in this exercise would depend on how much exposure time is allotted to each filter, the NS luminosity, and so on, a careful study of which we leave to future work.

\subsection{Implications for PSR J2144$-$3933 and similar pulsars}
\label{subsec:J2144}

Let us recollect that PSR J2144$-$3933, with spin-down age $t_{\rm NS} = 3.3 \times 10^8$ yr, is the coldest neutron star observed with a temperature upper limit of $\Tinf <$ 33000~K as obtained in Ref.~\cite{coldestNSHST}.
From Fig.~\ref{fig:dmax}, the $d_{\rm max}$ sensitivities of JWST-NIRCam, ELT-MICADO and TMT-IRIS in the F070W, $I$ and $Z$ filters to a $\Tinf = 20000$~Kelvin NS for an exposure time of $10^5$ seconds are, respectively, 170~pc, 210~pc, and 340~pc.
These are comparable to or greater than the parallax distance of PSR J2144$-$3933, 157--192 pc, which is the 1$\sigma$ range obtained after correcting for the Lutz-Kelker bias in radio parallax measurements~\cite{Deller2009:J2144dist,Verbiest2010LutzKelkerPulsar}.
This implies that a stronger upper limit on the temperature of this pulsar (or a potential measurement) is within the sensitivity of these filters.

In Table~\ref{tab:J2144-3933} we show the SNR = 5 ``discovery" reach and SNR = 2 ``upper limit" reach in the effective temperature $T_\infty$ of PSR J2144$-$3933 for these filters with exposure times of $10^4$~s and $10^5$~s, and for distances of 157 pc, 172 pc, 192 pc (the central value and the 1$\sigma$ edges of the parallax distance measurement). 
We see that in the best case, the future TMT-IRIS in the $Z$ band can reach down to 13350 K at SNR = 5 and 8510 K at SNR = 2, a significant improvement over the current upper bound.
Pressingly, today's JWST-NIRCam already has the sensitivity to set an SNR = 2 upper bound of 17880--23350 Kelvin, an improvement over the 27000--36000 Kelvin range of upper bounds set in Ref.~\cite{coldestNSHST} by varying the pulsar distance and radius.
{\bf This is one of the main results of our study.}

We note in passing that PSR J2144$-$3933 is not only the coldest NS observed, but is also the pulsar with the smallest known spin-down power~\cite{Deller2009:J2144dist}, and is notable for lying far beyond the so-called pulsar death line on the $P$-$\dot P$ plane, calling into question current models of pulsar radio emission~\cite{PulsarDeathLineAnomaly:PSRJ0250+5854}.
Any opportunity for closer scrutiny of this maverick pulsar should therefore be seized.
Observations of anomalous reheating could potentially lead to new astrophysical insights that explain the unusual properties of such slowly rotating radio pulsars.

While we have shown the imminent reaches of the temperature of PSR J2144$-$3933, other middle-aged pulsars with a measured upper bound on their temperatures are also ripe for observation at JWST, ELT and TMT, albeit with lower sensitivities. 
One such is PSR J1932+1059 (a.k.a. PSR B1929+10), with an upper limit on its effective temperature $T_\infty \lsim 5\times 10^5$~K~\cite{coolingcatalogue:Potekhin:2020ttj,coolingcataloguePotekhin:website}.
For its spin-down age of $3.1\times 10^6$~yr, one expects from minimal cooling models the effective temperature to be about $2\times10^4$~K~\cite{coolinganalytic:Ofengeim:2017cum,NSvIR:clumps2021}.
As the distance to this pulsar is $310\substack{+90 \\ -60}$ pc~\cite{Verbiest2012:pulsardists}, we see from Fig.~\ref{fig:dmax} that its passive-cooling temperature -- or a slightly reheated temperature -- may be cornered at JWST and ELT with exposure times of $10^6$~s and at TMT with $\lsim 10^5$~s.
Of course, a reheated temperature in excess of 40000 Kelvin (via a mechanism that is non-universal to pulsars so as to satisfy the upper limit on PSR J2144$-$3933's temperature) could be unveiled at these telescopes with much shorter exposure times.

\subsection{Potential neutron star candidates and pulsar catalogues}
\label{subsec:pulsarcats}

Although about $10^5$ NSs lie in the solar neighborhood of the Galaxy, fewer than 4000 have been found (see Sec.~\ref{sec:disc}), and among these fewer than 100 have had their luminosities measured~\cite{Potekhin:2013qqa}.
On the other hand, the ATNF catalogue lists about 280 pulsars within a distance of kpc as derived from dispersion measure data and the YMW16 model of electron column densities: see Sec.~\ref{sec:disc} for a discussion.
Thus hundreds, if not thousands, of discovered NSs wait for their temperature to be taken.

In addition, the recently instated radio telescope FAST in Guizhou, China has listed, at the time of writing, 825 newly found pulsars [188 in the Commensal Radio Astronomy FAST Survey (CRAFTS)~\cite{pulsarcatalogueFASTCRAFTS188} and 637 in the Galactic Plane Pulsar Snapshot (GPPS) survey v3.2.0~\cite{pulsarcatalogueFASTGPPS637,*pulsarcatFASTGPPS1Han:2021ekd,*pulsarcatFASTGPPS2Zhou:2023nns,*pulsarcatFASTGPPS3Su:2023vcw}], and CHIME in British Columbia, Canada has listed 25 pulsars~\cite{pulsarcatalogueCHIME25} (of which 14 are rotating radio transients~\cite{pulsarcatalogueCHIME2ndSet21}). 
From these catalogues, we have taken the information on these pulsars' dispersion measures and equatorial co-ordinates, and used the online tool in Ref.~\cite{distcalcPyGEDMOnline} (that is based on {\tt PyGEDM}~\cite{PriceElectronModelCompare:2021gzo}) to extract their distances from Earth as per the YMW16~\cite{YMW16:2017} and NE2001~\cite{NE2001:2002} models of Galactic electron densities.
We have tabulated these distances in Appendix~\ref{app:pulsarcats} (Tables~\ref{tab:pulsarcatFAST1},~\ref{tab:pulsarcatFAST2},~\ref{tab:pulsarcatCHIME}), in ascending order of the minimum of the two distances obtained from either model.
Our exercise shows that there are potentially 24 pulsars in the FAST-CRAFTS catalogue, 7 in the FAST-GPPS catalogue, and 5 in the CHIME catalogue that are within a kpc of Earth, for which luminosity measurements can be performed at JWST, EMT, and TLT with not-unreasonable exposure times. 
With continued observation, the rate of change of spin period $\dot P$ of these pulsars may be measured, from which their spin-down age $P/2\dot P$ inferred.
Combining the luminosity and age measurements will inform us about late-time reheating in these NSs. 
It must be noted that the distances we have inferred from the dispersion measure are far from certain, a point to which we turn in the next section.

\section{Discussion}
\label{sec:disc}

In this study we have estimated the maximum distance to neutron stars that JWST, ELT and TMT can observe with viable exposure times for stellar temperatures ranging between 2000--40000 Kelvin, which would make them the coldest NSs detected.
One challenge to note here is that the measurement of NS distances is fraught with large uncertainties.
These come particularly in the form of disagreement between methods of radio parallax and dispersion measure, and among inferences of dispersion measure distances from various Galactic electron distribution models. 
To illustrate with an example, the parallax estimate of PSR J2144$-$3933 (discussed in Sec.~\ref{subsec:J2144}), 157$-$192~pc, is consistent with the dispersion measure method in the TC93 model giving 180 pc but not in the NE2001 model giving 264 pc~\cite{Deller2009:J2144dist} and YMW16 model giving 289 pc~\cite{ATNF:2004bp}.
An even more egregious example is PSR J1057-5226, listed in the ATNF catalogue as the closest pulsar at 93 pc via the YMW16 model,
but is put at 1530~pc and $730 \pm 150$~pc by the TC93 and NE2001 models respectively~\cite{ClosestATNFDistsPosselt:2015bra}, and at 350$\pm$150~pc by a simultaneous analysis of the optical and X-ray spectrum~\cite{ClosestATNFOpticalXRay2010}; a parallax estimate of its distance is lacking to our knowledge.
Due to the highly uncertain and mutually inconsistent models of electron column densities~\cite{PriceElectronModelCompare:2021gzo}, dispersion measures yield less accurate distance measurements than radio parallax data, when available.
Such are the dispersion measure distance systematics that the YMW16 model can only guarantee that the uncertainty on 95\% of its estimates is less than 90\%, in itself a 50\% improvement over the NE2001 model (if the pulsar sample size of NE2001 were artificially increased to that of YMW16, that is)~\cite{YMW16:2017}.
As a reminder, typical uncertainties in parallax distance measurements are 10\%$-$20\%.
Refinement of dispersion measure methods for nearby pulsars warrants further study~\cite{BMRST}.
In any case, as discussed in Sec.~\ref{subsec:trends} the distance uncertainty may be eliminated from the temperature measurement if an NS' thermally emitted photons are observed in two or more filters with sufficient statistics.

As indicated by the right-hand y-axis of Fig.~\ref{fig:dmax}, there lurk a few million NSs within a kpc distance, yet only about 3500 pulsars have been discovered in all so far~\cite{ATNF:2004bp}.
While this is partly due to the limitation in sensitivity of radio telescopes, it is also likely due to suppression in the pulsing mechanism itself. 
It is generally believed that as pulsars get older than $10^7$ yr their rotational energy gets too weak to generate the pulsar beam.
The pulsar is then said to cross the ``pulsar death line", a.k.a. ``death valley", as drawn on a $P$-$\dot P$ diagram.
However, numerous pulsars (including PSR J2144-3933, as mentioned in Sec.~\ref{subsec:J2144}) have been recently observed to lie beyond different models of the death line~\cite{PulsarDeathLineAnomaly:PSRJ0250+5854,PulsarDeathLineAnomaly:PSRJ0901-4046,PulsarDeathLineAnomaly:PSRJ2144-3933,PulsarDeathLineAnomaly:PSRJ2251-3711}.
Further, around $10^5$ NSs in the Milky Way are estimated to similarly lie beyond the death line~\cite{PulsarDeathLineAnomaly:PSRJ2144-3933}.
These anomalies underline the fact that the precise mechanism generating pulsar beams is as yet poorly established.
They also suggest that with better radio sensitivity, the fraction of NSs discoverable as pulsars would be larger than previously supposed.

If late-stage reheating mechanisms such as those outlined in Appendix~\ref{app:reheat} were in place, one other way to discover large populations of neutron stars would be sky surveys.
In the near future, Rubin/LSST~\cite{Rubin1,Rubin2} that is designed to achieve a sensitivity of g-band AB magnitude 27.5 (corresponding to 36 nJy) will be able to detect NSs with effective temperatures a few times 10$^4$~Kelvin that are about 100 pc away~\cite{NSvIR:clumps2021}.
This means that there are potentially a few hundred $\Oc(10^4)$~Kelvin-hot NSs that may be discovered by sky surveys with Rubin/LSST, which would be a spectacular discovery. 
In the future, infrared surveys such as Roman/WFIRST~\cite{RomanWFIRST} may also uncover fainter neutron stars in the solar vicinity.
Getting such observational handles on reheating mechanisms gains significance when we consider that NSs have already been discovered that are hotter than expected for their age, notably the millisecond pulsars PSR J0437$-$4715~\cite{NStemp:J0437-4715:2003eb,NStemp:J0437-4715} and
PSR J2124$-$3358~\cite{NStemp:J2124-3358},
and the regular pulsars PSR J0108$-$1431~\cite{NStemp:J0108-1431,*NStemp2:J0108-1431:HST} and PSR B0950$+$08~\cite{NStemp:B0950+08}.
These and other similarly over-warm pulsars are consistent with rotochemical and vortex creep heating mechanisms; see Appendix~\ref{app:reheat}.

As mentioned in the Introduction, infrared imaging instruments may shed light on (after taking it from) poorly understood {\em central compact objects}, sources of steady X-rays with luminosities larger than their spin-down power.
CCOs are also marked by their relatively small magnetic field strengths ($\Oc(10^{10-11})$ Gauss), for which one explanation could be that an expected $10^{12-14}$ Gauss magnetic field of an NS at birth was buried in its crust by supernova debris falling back. 
This fallback could form around the NS a disk that could emit in the infrared, detection of which would favour models predicting the re-surfacing of the (small) magnetic field, and explain the non-detection of later-stage versions of young CCOs~\cite{Gotthelf2013CCOG}.
The infrared luminosity of the disk, if heated by the CCO's thermal X-rays, could be proportional to the X-ray luminosity, and the infrared spectrum could be power-law or blackbody.
Non-thermal emission from the NS could also result in a power law spectrum.
A recent search for these possibilities using NIRCam at JWST resulted in non-detection~\cite{CasAJWSTMilisavljevic:2024mbg}, however there are 9 confirmed and 3 suspected CCOs~\cite{CatalogueCCODeLuca} for which analogous measurements may be taken at JWST, ELT, and TMT.

Since their discovery, neutron stars have served as excellent testing grounds for fundamental physics. 
Today's and tomorrow's telescopes now have the opportunity to be pioneering thermometers, with the potential to make transformative discoveries.

\section*{Acknowledgments}

We thank Joseph Bramante,
Raghuveer Garani,
Pranav Kalsi,
Avinash Kumar Paladi, and
Akshank Tyagi for discussion, 
Jasnoor Singh for a script to extract distances to pulsars in the FAST catalogues,
and Koichi Hamaguchi and Natsumi Nagata for comments on the manuscript.

\appendix

\renewcommand{\arraystretch}{.97}
\begin{table*}[t]
    \centering
    \begin{tabular}{|l|c|c|c|}
    \hline
        \textbf{pulsar} & \textbf{dispersion measure} (pc/cm$^3$) & \textbf{distance YMW16} (pc) & \textbf{distance NE2001} (pc) \\ \hline
        J0745-0411 & 10 & \textcolor{blue}{\bf 149.2} & {\bf 670.3} \\ 
        J0623+0051 & 20 & \textcolor{blue}{\bf 153.3} & 1073.7 \\ 
        J1720-0533 & 37.8 & \textcolor{blue}{\bf 191.4} & 1380.1 \\ 
        J1721-0854 & 49.1 & \textcolor{blue}{\bf 201.0} & 1616.0 \\ 
        J1708-0424 & 49.3 & \textcolor{blue}{\bf 230.3} & 1858.7 \\ 
        J1712-1124 & 69.3 & \textcolor{blue}{\bf 232.5} & 2253.3 \\ 
        J1020+4056 & 9.9 & 1050.5 & \textcolor{blue}{\bf 418.0} \\ 
        J1115-0958 & 9.9 & \textcolor{blue}{\bf 506.3} & \textcolor{blue}{\bf 467.1} \\ 
        J0844-0329 & 15 & \textcolor{blue}{\bf 516.1} & {\bf 815.8} \\ 
        J1751-0542 & 64 & \textcolor{blue}{\bf 522.3} & 1950.9 \\ 
        J2355+0049 & 11 & {\bf 943.8} & \textcolor{blue}{\bf 545.7} \\ 
        J0846-1049 & 13.7 & \textcolor{blue}{\bf 562.4} & {\bf 866.3} \\ 
        J1737-0514 & 78 & \textcolor{blue}{\bf 568.1} & 2525.5 \\ 
        J0631+4147 & 24.5 & \textcolor{blue}{\bf 574.0} & {\bf 823.7} \\ 
        J2230-0336 & 10.3 & {\bf 851.7} & \textcolor{blue}{\bf 587.5} \\ 
        J0756-0517 & 21.7 & {\bf 673.6} & 1272.2 \\ 
        J0941+4542 & 17.6 & 1794.6 & {\bf 680.5} \\ 
        J0203-0150 & 19.2 & 1841.9 & {\bf 772.2} \\ 
        J0803-0937 & 21.5 & {\bf 840.0} & 1336.5 \\ 
        J0521+5647 & 21.8 & {\bf 982.9} & {\bf 863.3} \\ 
        J1742-0559 & 96.2 & {\bf 910.1} & 3035.3 \\ 
        J0924+6102 & 21.8 & 1989.0 & {\bf 919.2} \\ 
        J0447+2446 & 32.8 & {\bf 920.5} & 1066.6 \\ 
        J1611-0113 & 19.5 & 1183.7 & {\bf 990.2} \\ 
        \hline
 \end{tabular}
    \caption{A subset of the 188 pulsars in the publicly available FAST-CRAFTS catalogue~\cite{pulsarcatalogueFASTCRAFTS188}, 
    along with dispersion measure distances derived in this work using the pulsars' equatorial co-ordinates (indicated in their names) and the YMW16 and NE2001 electron column density models.
    The pulsars are listed in ascending order of the minimum of the distance obtained in either model.
    Only pulsars with this minimum smaller than 1 kpc are listed here, while the rest of the catalogue is uploaded to arXiv as an ancillary file.
    Distances smaller than 600 pc~(1 kpc) are highlighted in \textcolor{blue}{\bf blue} ({\bf bold}). 
   The uncertainties on most of the YMW16 distances are estimated to be $<$90\%, but for some it could be larger in both models.
     See Secs.~\ref{subsec:pulsarcats} and \ref{sec:disc} for further details.}
     \label{tab:pulsarcatFAST1}
\end{table*}

\renewcommand{\arraystretch}{.97}
\begin{table*}[t]
    \centering
    \begin{tabular}{|l|c|c|c|}
    \hline
        \textbf{pulsar} & \textbf{dispersion measure} (pc/cm$^3$) & \textbf{distance YMW16} (pc) & \textbf{distance NE2001} (pc) \\ \hline
      J0653+0443g & 27.3 & \textcolor{blue}{\bf 181.3} & 1251.1 \\ 
        J0623+0220g & 32.1 & \textcolor{blue}{\bf 437.5} & 1399.5 \\ 
        J1908+1035g & 10.9 & {\bf 670.9} & \textcolor{blue}{\bf 593.4} \\ 
        J1854+0704g & 10.8 & {\bf 647.2} & {\bf 973.3} \\ 
        J2011+3006g & 14 & {\bf 943.9} & 1225.0 \\ 
        J1844+0315g & 23 & {\bf 974.2} & 1334.9 \\ 
        J1857+0642g & 21.6 & {\bf 993.5} & 1586.7 \\ 
        \hline
 \end{tabular}
    \caption{Same as Table~\ref{tab:pulsarcatFAST1} but a subset of the 637 pulsars in the publicly available FAST-GPPS catalogue~\cite{pulsarcatalogueFASTGPPS637,*pulsarcatFASTGPPS1Han:2021ekd,*pulsarcatFASTGPPS2Zhou:2023nns,*pulsarcatFASTGPPS3Su:2023vcw}.
    The rest of the catalogue is uploaded to arXiv as an ancillary file.
    FAST adds the suffix ``g" to these pulsar names to indicate a position uncertainty of about 1.5$'$, which may propagate into the dispersion measure distance estimate.}
     \label{tab:pulsarcatFAST2}
\end{table*}

\begin{table*}[t]
    \centering
    \begin{tabular}{|l|c|c|c|}
    \hline
        \textbf{pulsar} & \textbf{dispersion measure} (pc/cm$^3$) &  \textbf{distance YMW16} (pc) & \textbf{distance NE2001} (pc) \\ \hline
        J0746+55 & 10.5 & \textcolor{blue}{\bf 410.4} & \textcolor{blue}{\bf 468.5} \\ 
        J1105+02 & 16.5 & 1579.6 & {\bf 690.6} \\ 
        J0854+54 & 17.8 & 1330.7 & {\bf 710.3} \\ 
        J0658+29 & 40.05 & {\bf 891.8} & 1350.5 \\ 
        J1252+53 & 20.7 & 3190.6 & {\bf 995.7} \\
        J1130+09 & 21.9 & 5184.6 & 1023.5 \\ 
        J1541+47 & 19.4 & 2033.8 & 1037.8 \\ 
        J2215+4524 & 18.5 & 1131.2 & 1579.2 \\ 
        J0741+17 & 44.3 & 1206.8 & 1711.1 \\ 
        J2355+1523 & 26 & 3422.3 & 1362.1 \\ 
        J1838+5051 & 21.8 & 1702.6 & 1539.6 \\ 
        J0227+33 & 27.56 & 2050.7 & 1741.7 \\ 
        J0209+58 & 56 & 1751.2 & 2142.3 \\ 
        J0653-06 & 83.7 & 2140.3 & 2984.7 \\
        J2138+69 & 46.6 & 2486 & 2204.4 \\ 
        J2237+2828 & 38.1 & 3989.9 & 2279.5 \\ 
        J0121+5329 & 87.35 & 2529.2 & 3213.6 \\ 
        J2116+37 & 44 & 2914.7 & 2803.5 \\ 
        J2208+46 & 63 & 3148.4 & 3095.9 \\ 
        J2108+45 & 84 & 3592.3 & 3778.3 \\ 
        J1943+58 & 71.2 & 7874.5 & 4722.8 \\ 
        J2057+46 & 218.962 & 4939.3 & 7105 \\ 
        J0012+5431 & 131.3 & 5425 & 5727.3 \\ 
        J2008+3758 & 143 & 5923.7 & 5825.4 \\ 
        J2005+38 & 192.8 & 7647 & 7633.1 \\ \hline
    \end{tabular}
    \caption{The 25 pulsars (including 14 rotating radio transients) in the CHIME catalogue~\cite{pulsarcatalogueCHIME25} along with dispersion measure distances derived in this work using the pulsars' equatorial co-ordinates (indicated in their names) and the YMW16 and NE2001 electron column density models. 
       The pulsars are listed in ascending order of the minimum of the distance obtained in either model. 
     Distances smaller than 600 pc~(1 kpc) are highlighted in \textcolor{blue}{\bf blue} ({\bf bold}).
     The uncertainties on most of the YMW16 distances are estimated to be $<$90\%, but for some it could be larger in both models.
          See Secs.~\ref{subsec:pulsarcats} and \ref{sec:disc} for further details.}
     \label{tab:pulsarcatCHIME}
\end{table*}

\section{Reheating mechanisms}
\label{app:reheat}

In this appendix we briefly review neutron star late-stage reheating mechanisms that arise in plausible astrophysical settings as well as via more exotic agents such as a hidden sector of states and dark matter. 
It must be remembered that none of these mechanisms have been unambiguously observed, nor are they soundly established in theory.
Non-observation of NS luminosities would place upper limits on all their strengths.

\subsection{Astrophysical mechanisms}

{\bf Rotochemical heating}~\cite{Rotochem:FernandezReisenegger:2005cg,*Rotochem:Reisenegger:2006ky,*Rotochem:PetrovichReis:2009yh,*Rotochem:GonzalezJimenezReis:2014iia,*Rotochem:Gusakov2021:Refined}.
As an NS spins down, the decreasing centrifugal forces drive NS material out of beta equilibrium.
Processes restoring the balance in chemical potentials would then deposit heat.
It is for certain nucleon pairing models of superfluidity in NSs with $\lsim \Oc(10)$~ms natal periods that this mechanism will take effect~\cite{NSvIR:Hamaguchi:RotochemicalPure2019,NSvIR:Hamaguchi:RotochemicalvDM2019}, which must be contrasted against findings that initial spin periods are likely $\Oc(10-100)$~ms~\cite{NatalPeriod:Faucher-Giguere:2005dxp,*NatalPeriod:2010,*NatalPeriod:2012,*NatalPeriod:2013,*NatalPeriod:Igoshev:2013yua,*NatalPeriod:Gullon:2014dva,*NatalPeriod:Gullon:2015zca,*NatalPerod:Muller:2018utr}, with the caveat that NS initial periods depend on the details of the supernova event. 
Rotochemical heating has been invoked to explain the temperatures of pulsars that are otherwise too high for their spin-down age, and is also shown to satisfy the upper bound on the effective temperature of PSR J2144$-$3933~\cite{NSvIR:Hamaguchi:RotochemicalPure2019}.
One variant of this mechanism is reheating due to upsetting of nuclear statistical equilibrium in the NS crust as opposed to chemical equilibrium in the core~\cite{Rotochem:GusakovReisenegger2015:CrustNSE}.

{\bf Vortex creep}~\cite{VortexCreep:Anderson:1975zze,*VortexCreep:1984,*VortexCreep:1989,*VortexCreep:1993,*VortexCreep:VanRiper:1994vp,*VortexCreep:Larson:1998it}.
Nucleon-superfluid vortex lines in the inner crust may be pinned to the background lattice of nuclei.
In addition to this pinning force, vortex lines experience a Magnus force due to 
differences in rotational speed between the normal and superfluid matter, which causes them to move outward.
This creep motion induces friction, which then dissipates energy.
The heating luminosity of this mechanism is proportional to the angular deceleration $\dot \Omega$, and the proportionality constant is expected to be the same for all NSs.
Intriguingly, the proportionality constant reconstructed from observation assuming thermal equilibrium,  $J_{\rm obs} \equiv 4\pi R_{\infty}^2 \sigma_{\rm SB} T^4_\infty/|\dot \Omega| $, is indeed found to be roughly uniform across an ensemble of millisecond and ordinary pulsars that are too hot for their spin-down age~\cite{VortexCreepPure:Hamaguchi:2023tmr}.
The fitted range $J \simeq 10^{35.9-36.8}$~Joule sec is also consistent with theoretical predictions in a mesoscopic approach of estimating the vortex pinning force.

{\bf Magnetic field decay}~\cite{BfieldDecay:1992Reisenegger}.
Through such processes as ohmic decay, ambipolar diffusion and Hall drift, the magnetic field of the NS could dissipate energy into the stellar material.
Rough estimates of the heating luminosity~\cite{NsvIR:otherinternalheatings:Reisenegger,NSvIR:otherinternalheatings:2022} indicate that old isolated NSs have magnetic field strengths in the right range to produce $T_\infty \lsim 10^4$~K.  

{\bf Crust cracking}~\cite{CrustCrack:1971BaymPines}. 
As the NS spins down, the ellipsoidal crust gradually becomes spherical, acquires stress, and breaks, liberating the energy of the accumulated strain.
Crust cracking can result in reheating temperatures of $T_\infty \simeq 10^4$~K~\cite{NsvIR:otherinternalheatings:Reisenegger}.

We remark that accretion of interstellar material (ISM) is unlikely to reheat old NSs to detectable luminosities.
Around old, isolated NSs with periods $< 10^3$ sec, the NS magnetic field drives a ``pulsar wind" of ISM outflow that is much denser than inflowing material.
Thus kinetic pressure prevents accretion, and the NS is said to be in an ejector phase~\cite{NSvIR:ISMaccretionphases:Treves:1999ne}.
Were the ISM to overcome a weak pulsar wind, its accretion  would still be thwarted by the centrifugal acceleration provided by the co-rotating magnetosphere, which would disperse the ISM. 
The NS is then said to be in the {\em propellor phase}.
For unusually slow-rotating NSs with periods $> 1000$ seconds, infalling ISM would travel along the NS' magnetic field lines so that only a small polar region accretes, which may be discriminated from thermal emission.

\subsection{Non-standard mechanisms}

{\bf Dark matter capture}~\cite{BramanteRajCompactDark:2023djs,snowmass:ExtremeBaryakhtar:2022hbu,*snowmass:Carney:2022gse}.
Perhaps the most minimal theoretical setup to reheat old, isolated NSs, this mechanism suggests that the incident energy flux of dark matter particles ambient to a neutron star in the solar vicinity has a heating luminosity corresponding to blackbody temperatures of about 2000~K.
The steep gravitational potential of NSs (with surface escape speeds $\gsim 0.5 c$) serves to both enhance gravitational focusing of the dark matter flux and accelerate the dark matter to a kinetic energy comparable to its mass energy.
Via kinetic energy transfer~\cite{NSvIR:Baryakhtar:DKHNS}, scattering of the dark matter on the NS' nucleons, electrons, muons, possible hyperons, and even crustal components like nuclear pasta and superfluid neutrons can impart heat~\cite{NSvIR:Raj:DKHNSOps,NSvIR:Bell2018:Inelastic,NSvIR:Bell2019:Leptophilic,NSvIR:Riverside:LeptophilicShort,NSvIR:Riverside:Leptophiliclong,NSvIR:Bell:ImprovedLepton,NSvIR:GaraniHeeck:Muophilic,NSvIR:SelfIntDM,NSvIR:Hamaguchi:RotochemicalvDM2019,NSvIR:GaraniGenoliniHambye,NSvIR:Queiroz:Spectroscopy,NSvIR:Bell:Improved,NSvIR:Bell2020improved,NSvIR:anzuiniBell2021improved,NSvIR:Marfatia:DarkBaryon,NSvIR:DasguptaGuptaRay:LightMed,NSvIR:zeng2021PNGBDM,NSvIR:Queiroz:BosonDM,NSvIR:HamaguchiEWmultiplet:2022uiq,NsvIR:HamaguchiMug-2:2022wpz,NSvIR:PseudoscaTRIUMF:2022eav,NSvIR:InelasticJoglekarYu:2023fjj,NSvIR:Hamaguchi:VortexCreepvDM2023}.
In addition, in some models captured dark matter can thermalize~\cite{Bertoni:2013bsa,NSvIR:GaraniGenoliniHambye,NSvIR:BellThermalize:2023ysh} with the NS and self-annihilate or co-annihilate with nucleons to deposit more heat, helping observation prospects~\cite{Kouvaris:2007ay,deLavallaz:2010wp,NSvIR:Baryakhtar:DKHNS,NSvIR:Raj:DKHNSOps,NSvIR:Pasta,NSvIR:IISc2022,NSvIR:hylogenenesis1:2010,NSvIR:hylogenenesis2:2011,NSvIR:coann-nucleons:JinGao2018moh,NSvIR:Marfatia:DarkBaryon,DMann:bubblenucleation:Silk2019,NSvIR:magneticBH:2020}.
JWST may also reveal the overheating of exoplanets by self-annihilations of captured dark matter~\cite{Leane:2020wob}.

Higher temperatures up to 40000~K can be obtained via Bondi accretion of strongly-interacting dark matter particles with dissipative fluid behavior in clumps/microhalos that periodically encounter NSs~\cite{NSvIR:clumps2021}.
In the absence of Bondi accretion, these clumps could temporarily impart temperatures $> 10^{3-4}$~K to NSs, which can be observed in a sky survey looking for a fraction of overheated NSs in an ensemble.
These temperatures may also be achieved by a combination of tidal heating and kinetic energy transfer by dark matter states that experience a new long-range force with baryons, with the heating energy sourced mainly by the long-range potential that effectively enhances the gravitational potential of the neutron star~\cite{NSvIR:tidalfifthforce:Gresham2022}.

{\bf Nucleon Auger effect}~\cite{NSheat:DarkBary:McKeen:2020oyr,*NSheat:Mirror:McKeen:2021jbh}.
Nucleons may disappear from NSs through scattering and decay processes, notably to final states involving dark sector states carrying baryon number, ``dark baryons".
The resulting hole in the Fermi sea gets filled rapidly by surrounding nucleons with higher energy, accompanied by spillage of electromagnetic and kinetic energy, reminiscent of the Auger effect in semi-conductors~\cite{NSheat:DarkBary:McKeen:2020oyr,*NSheat:Mirror:McKeen:2021jbh,NSheat:Mirror:Goldman:2022brt,*NSheat:Mirror:Goldman:2022rth,*NSheat:Mirror:Berezhiani:2020zck}.
The heat liberated in this process can raise NS temperatures to 40000~K.

{\bf Baryon number-violating decay}~\cite{Davoudiasl:2023peu}.
The decay of the neutron to a pion and positron via an ultra-light scalar mediator is better constrained by neutron star overheating due to neutron mass energy deposition than by terrestrial experiments like Super-Kamiokande.
This process too can impart NS temperatures of 40000~K.

\section{Pulsar catalogues}
\label{app:pulsarcats}

In this Appendix we tabulate pulsars from the catalogues of FAST~\cite{pulsarcatalogueFASTCRAFTS188,pulsarcatalogueFASTGPPS637,*pulsarcatFASTGPPS1Han:2021ekd,*pulsarcatFASTGPPS2Zhou:2023nns,*pulsarcatFASTGPPS3Su:2023vcw}, and CHIME~\cite{pulsarcatalogueCHIME25} and their dispersion measure distances we obtained using the online tool in Ref.~\cite{distcalcPyGEDMOnline}, as discussed in Sec.~\ref{subsec:pulsarcats}.

\bibliography{refs}

\end{document}